# Soft X-ray irradiance measured by the Solar Aspect Monitor on the Solar Dynamic Observatory Extreme ultraviolet Variability Experiment

C. Y. Lin[1,2]

S. M. Bailey[2]

A. Jones[3]

D. Woodraska[3]

A. Caspi[4]

T. N. Woods[3]

F. G. Eparvier[3]

S. R. Wieman[5]

L. V. Didkovsky[5]

[1] Department of Physics, University of Texas at Arlington, Arlington, Texas, USA

[2] Center for Space Science and Engineering Research, Virginia Tech, Blacksburg, Virginia, USA

[3] Laboratory for Atmospheric and Space Physics, Boulder, Colorado, USA

[4] Southwest Research Institute, Boulder, Colorado, USA

[5] Space Sciences Center, University of Southern California, Los Angeles, California, USA

Corresponding author: incen@vt.edu

Key Points:

1. Methods of removing particle contamination are proposed and analyzed.

2. Broadband soft X-ray irradiance is derived and validated.

3. SAM is capable of resolving spatial and spectral irradiance.

## Abstract

The Solar Aspect Monitor (SAM) is a pinhole camera on the Extreme-ultraviolet Variability Experiment (EVE) aboard the Solar Dynamics Observatory (SDO). SAM projects the solar disk onto the CCD through a metallic filter designed to allow only solar photons shortward of 7 nm to pass. Contamination from energetic particles and out-of-band irradiance is, however, significant in the SAM observations. We present a technique for isolating the 0.01–7 nm integrated irradiance from the SAM signal to produce the first results of broadband irradiance for the time period from May 2010 to May 2014. The results of this analysis agree with a similar data product from EVE's EUV SpectroPhotometer (ESP) to within 25%. We compare our results with measurements from the Student Nitric Oxide Explorer (SNOE) Solar X-ray Photometer (SXP) and the Thermosphere Ionosphere Mesosphere Energetics and Dynamics (TIMED) Solar EUV Experiment (SEE) at similar levels of solar activity. We show that the full-disk SAM broadband results compare well to the other measurements of the 0.01–7 nm irradiance. We also explore SAM's capability toward resolving spatial contribution from regions of solar disk in irradiance and demonstrate this feature with a case study of several strong flares that erupted from active regions on March 11, 2011.

# 1 Introduction

Solar radiation from about 0.1 to 20 nm, or soft X-ray (SXR) irradiance, [Barth et al., 1988] is mostly absorbed at altitudes between 100 and 150 km and is the major driver of both the neutral and ionized atmosphere [Fuller-Rowell et al., 2004]. Solar radiation at these wavelengths, is critical to electron densities in the ionospheric E-region [Sojka et al., 2006; Solomon, 2006]. Changes in neutral density driven by SXR variability impact satellite drag and must be considered for orbit prediction [Emmert et al., 2010]. The important role solar SXR irradiance also plays in the nitric oxide production was recognized in the late 1980s [Barth et al., 1988; Siskind et al., 1990].

The history of solar SXR observation can be traced back to the l960s. Sounding rocket measurements led to several satellite-based experiments, including SOLRAD [Dere et al., 1974], Orbiting Solar Observatory (OSO) [Hall and Hinteregger, 1970; Hall, 1971], the Atmospheric Explorer (AE) satellites [Gibson and Van Allen, 1970; Hinteregger et al., 1981], SkyLab (1973 – 1979) [Vaiana et al., 1976], and Solar Maximum Monitor (SMM) (1980) [Acton et al., 1980; Bohlin et al., 1980]. The Soft X-ray Telescope (SXT) on Yohkoh took full-disk images of the Sun between 0.2–3 nm from 1991 to 2001 [Ogawara et al., 1992]. SXR irradiance was estimated with an isothermal spectral model and modeled by means of differential emission measures (DEM) [Acton et al., 1999]. The Soft X-ray Photometer (SXP) on the Student Nitric Oxide Explorer (SNOE) had several broadband channels making daily measurements of SXR irradiance with bandpasses of 2–7 nm, 6–19 nm, and 17–20 nm [Bailey et al., 2000; 2006]. Solar Extreme ultraviolet Monitor (SEM) on the SOlar and Heliospheric Observatory (SOHO) has continuously measured the solar extreme ultraviolet (EUV) irradiance since 1996 [Judge et al., 1998; Ogawa et al., 1998; Wieman et al., 2014]. Its zeroth-order channel monitors the full-disk solar irradiance from 0.1 to 50 nm, and has a first-order channel measuring the He II 30.4 nm irradiance. The National Oceanic and Atmospheric Administration (NOAA) Geostationary Operational Environmental Satellite (GOES) series satellites carry X-Ray Sensor (XRS), measuring X-ray flux at 0.05–4 nm (XRS-A) and 0.1–0.8 nm (XRS-B) [Machol J. and R. Viereck, 2015]. The Solar EUV Experiment (SEE) onboard the Thermosphere, Ionosphere, and Mesosphere Energetics and Dynamics (TIMED) satellite carries the X-ray Ultraviolet (XUV) Photometer System (XPS) and the EUV Grating Spectrograph (EGS) to provide daily irradiance of XUV between 0.1 and 35 nm in broad bands and EUV spectra with 0.4 nm resolution between 27 and 200 nm, respectively [Woods et al., 1999, 2004]. The Solar Radiation and Climate Experiment (SORCE) XPS [Woods et al., 2008a] is nearly identical to the one on TIMED/SEE. The X-Ray Telescope (XRT) on Hinode started taking X-ray images in 2006 [Golub et al., 2007; Kano et al., 2008]. Woods et al. [2004] gave a detailed historical account of the spaceborne measurements of solar EUV and SXR irradiance. The difficulties in performing radiometric calibrations at these wavelengths, especially for the older instruments, and poor knowledge of the complicated solar spectrum underlying broadband irradiance measurements have yielded significant discrepancies among the various solar SXR data sets and models [Solomon and Qian, 2005]. Theoretical and empirical reference models [Dere et al., 1997; Tobiska et al., 2000; Warren, 2005; Richards et al., 2006; Chamberlin et al., 2007, 2008] of the solar spectrum have been established to better assist climate models; however, comparison of modeled atmospheric response using solar measurements with available atmospheric data has shown that the SXR may be underestimated in the solar reference models and sometimes needs to be scaled by a factor of 2 or more [Bailey et al., 2002]. This uncertainty is verified in studies of ion and electron densities as the production of these species is in proportion to the solar ionizing irradiance [Solomon, 2006]. In contrast, it was found that the SXR irradiance predicted by the XPS Level 4 model was overestimated by a factor of 4–8 [Caspi et al., 2015]. The Amptek X123-SDD on the sounding rocket measured much lower irradiances than what the XPS model predicted. It is concluded that the broadband models may over-predict or under-predict, depending on the situation. Therefore, actual measurements of SXR irradiances are crucial to resolve these issues.

Launched in 2010, Solar Dynamics Observatory (SDO) is the first mission of NASA's Living With a Star (LWS) program. The goal of SDO is to understand the solar variability and its impact to the terrestrial technology and society. SDO facilitates not only better knowledge of heliophysics by studying the evolution of solar magnetic fields and its atmosphere, but also the connection between solar activity and terrestrial effects [Pesnell et al., 2012]. The Extreme ultraviolet Variability Experiment (EVE) is one of three instruments on SDO [Hock et al, 2012; Woods et al., 2012]. EVE measures the SXR and EUV solar spectrum at 0.01–105 nm. The Solar Aspect Monitor (SAM) is a pinhole camera on EVE. It is designed to image the solar disk shortward of 7 nm (photon energy above 180 eV) every 10 seconds. Details of the SAM instrumentation are covered in Section 2. The EUV SpectroPhotometer (ESP) on EVE is an expanded version of SOHO/SEM. It is a non-focusing broadband spectrograph. Its quad-diode (QD) channel produces broadband SXR irradiance measurements in the same spectral band as SAM [Didkovsky et al., 2012].

Retrieving full-disk irradiance from the SAM images suffers an unanticipated difficulty due to the radiative environment. In this study, we propose methods to minimize the contamination, derive broadband irradiance at SXR wavelengths from the SAM images, and compare the results with the ESP zeroth-order QD irradiance product. Our goal is to show that SAM yields a reasonable full-disk estimate of the broadband SXR irradiance compared to ESP as the first step toward our eventual goal of obtaining spectral measurements from SAM. The latter will be explored in detail in a follow-on paper. This first step is important as there is currently no other data set that provides spectral irradiance at the similar bandwidth and has comparable spectral and temporal resolution to validate SAM once its spectrum unfolds to its full resolution. On the other hand, SAM's imaging also provides the capability to resolve incoming solar SXR flux spatially and spectrally with a 15-arcsecond resolution. This makes SAM a unique addition to the ESP, which provides higher cadence broadband irradiance, but at a spatial resolution of a quarter of the Sun, or the Atmospheric Imaging Assembly (AIA, also on the SDO), whose high-resolution cameras observing at a few EUV wavelengths between 9–35 nm [Lemen et al., 2012]. We demonstrate SAM's spatial capabilities by retrieving the localized contribution from active regions (ARs) during flares.

## 2 Instrument and Observations

SAM shares a CCD on EVE's Multiple EUV Grating Spectrograph (MEGS-A) [Woods et al., 2012; Hock et al., 2012] as shown in Figure 1. It operates in two modes. In SXR photon-counting mode, a Ti/Al/C foil filter is in position to allow photons shortward of 7 nm (>180 eV) to reach the detector and forms an SXR image onto the CCD. Dark measurements are made when the pinhole is covered by a dark filter. This dark mode is performed for about one minute per day during nominal operation. SAM operates in the SXR mode more than 23 hours per day.

The SAM entrance aperture is a 26 µm diameter pinhole set 32 cm from the CCD. The size of a CCD pixel is 15 µm by 15 µm, which gives approximately 15 arcsecond resolution of the solar disk. SAM projects the solar disk onto the corner of the CCD where spectral lines from MEGS-A are dim. Though few, some MEGS-A photons appear and contaminate the SAM images as discussed below. Figure 2 shows a contrast-enhanced one-hour integration of the SAM images with a clear limb and active regions [the LASP website]. Because the detector is tilted by 17° (accommodating the MEGS-A design) relative to the normal of the pinhole-detector axis, projection of the Sun is slightly elliptical with a semi major axis of about 215 pixels in the E/W direction and a minor axis of about 205 pixels in the N/S direction. Coronal loops have been seen to extend about 50 pixels above the limb surface. For the Si-based CCD used on EVE, an electron-hole pair is created when equivalent energy, $E_{si}$, of 3.63 eV impacts a CCD pixel. The number of electrons required to register one data number (DN) is defined as *gain*. Lab calibrations show that the gain of the SAM CCD, $G$, is 2.47 electrons per DN [Hock et al., 2012]. The conversion from DN to photon energy simply follows Equation 1.

$$E = E_{si} \cdot G \cdot DN_{corrected} \quad (1)$$

Dividing the constant, 1239.84, in eV·nm by the energy, $E$, in eV, yields wavelength in nm. Units of nm, eV, and DN are used interchangeably in the following context. Equation 1 should only be used for dark-count corrected DN values. The key values used in the data processing are referred to in Table 1. Two components contributing to the background dark counts are the electronic bias and the CCD thermal noise. Four virtual pixels are placed at the beginning of each CCD read-out row and are the indication of charge amplifier signal prior to reading actual CCD pixels. The average of these four pixels is obtained and subtracted off from all the pixels of interest in that row for every frame, and this provides the dark-corrected signal, $DN_{corrected}$. At the temperature at which the EVE CCDs operate, the thermal noise contribution is about 3 DN/s with uncertainty of about 2 DN/s. Both of these two terms are relatively small compared to the contamination from multiple background sources that appear on the SAM corner of the CCD.

As the goal is to obtain solar irradiance at SXR wavelengths, the SXR filter is designed to allow high-energy photons. The thickness of each material of the SAM foil filter is 80 nm for carbon (C), 320 nm for titanium (Ti), and 200 nm for aluminum (Al). The silicon (Si) effective layer in the CCD is 45 microns. With these parameters, the SAM transmission is modeled using tabulated atomic scattering factors [Henke et al., 1993] and shown in Figure 3. Sharp edges in the transmission are shaped by each material: C at around 4.5 nm, Ti from 1.5 to 3 nm, and Al at around 0.7 nm. The declining of transmission below 0.5 nm is defined by the silicon absorption of the detector. This response function is slightly different than the one in the work presented by Hock et al [2012] due to improved knowledge of the Si thickness. SAM's high cadence and low count rate approach brings challenges to distinguish between irradiance and contamination from energetic particles with similar energy levels in the data processing attempts. Study of the raw DN counts on the SAM images reveals the semi-diurnal encounter of the SDO spacecraft with the particles in the outer (electron) Van Allen radiation belt. Some of these particles strike the CCD directly and some cause Bremsstrahlung radiation from the supporting material around the CCD, resulting in high-DN read-outs on the CCD pixels. The SAM dark filter is in position about one minute per day. These 1-minute dark measurements are insufficient to directly determine the changing particle environment that the spacecraft encounters every 10 seconds. Therefore, a more involved method has to be performed to remove the particle contribution from the measurements.

SAM projects the solar image onto a relatively dark corner of the MEGS-A CCD, with this dark region being 512 pixels by 512 pixels and the full CCD size 2048 pixels by 1024 pixels. To include any possible flares occurring at the limb and the outer corona, the size of the sunlit image assumed for the SAM analysis is 320 pixels by 240 pixels. This area is referred to as *illuminated area* or *IA* in the following context. A narrow strip of non-sunlit area below the IA is chosen to represent the particle environment. It is located close enough to the IA so that contamination effects on this portion of the CCD can be assumed identical to non-solar sources on IA. We verify this assumption below. This area will be hereafter referred to as *unilluminated area* or *UA*. The size of the UA is chosen to be 480 pixels by 160 pixels so that there are the same number (67,200) of pixels on both types of images. These two areas are indicated by the white boxes in Figure 1. During nominal science mode, each complete MEGS-A CCD image provides a pair of SAM IA and UA images, from which SXR irradiance can be obtained using the techniques presented in this paper. Over 8,000 image pairs are produced in one day of nominal operations. Due to the low signal-to-contamination ratio, the entire solar disk is not readily apparent in a single image except in the active regions. The top panel of Figure 4 shows the typical DN histograms of the IA (red) and UA (black) images for a non-flaring day. When the Sun is quiet, the IA and UA distributions lie closely together and do not show apparent differences below ~20 DN and beyond 1000 DN. The non-zero UA distribution is indicative of non-photon sources at all DNs. It is expected that information of the incoming solar SXR flux lies in the difference of the IA and UA distribution. Hereafter, unless otherwise specified, the difference of the IA and UA histograms is simply referred as the *histogram*. The out-of-band MEGS-A photons with energy less than 130 eV equivalently contribute less than 15 DN on a pixel if all energy of a photon is assumed to be entirely deposited on one CCD pixel. By only taking pixels of values greater than 15 DN into account, most of the MEGS-A photons (energy < 130 eV) are assumed to be excluded in this broadband approach.

## 3 Channels

Conversion from broadband solar measurements to irradiance requires the instrument response function and a solar reference spectrum. Scaling factors are calculated to convert measured quantity (either voltage or current) to solar irradiance [Bailey et al., 2006]. Figure 3 shows the unit-less modeled transmission function of SAM. The Solar Irradiance Reference Spectra (SIRS) for the 2008 Whole Heliosphere Interval (WHI) [Chamberlin et al., 2009; Woods et al., 2008b], measured on 14 Apr 2008, has a spectral resolution of 0.5 nm, and is the reference spectrum implemented in this study. The broadband SAM measurements in DN is essentially the incoming solar irradiance weighted by the device response function.

$$DN = m \cdot f \cdot I_{sun} \qquad (2)$$

where $m$ is $3.68 \times 10^{-9}$ DN/(W·m$^{-2}$) and $f$ is a unit-less weighting factor and is the ratio of the integral of the weighted solar irradiance by the instrument response function to its non-weighted integral.

Information regarding solar irradiance and contamination of particles reside in the DN histogram as shown in Figure 4. In order to illustrate the differing information received by different portions of the histograms, we define fifteen DN channels. This approach will help isolate contributions from the solar photons and particles. The channels were originally defined by the energies of bright emission lines. We examined uniform digitization in equal-size DN bins, but found that the best approach is to use finer (coarser) resolution toward shorter (longer) wavelengths.

Therefore, the channels are defined based on DN and with non-uniform DN ranges. The bandwidth is 100 DN for channels including DN values between 100 and 1,000 and 1,000 DN for those including DN values above 1,000. Direct interpretation from the integral of a narrow DN range to narrowband irradiance creates significant uncertainties. These uncertainties can be minimized by taking the difference of integrals of two broader bands of different DN ranges whose difference is the bandwidth of the narrowband of interest As an example, channel #1 ($7000 \leq DN < 16382$) is the difference of the integrals of 15–16382 DN ($DN_{1,high}$) and that of 15–6999 DN ($DN_{1,low}$). Fifteen channels are tabulated in Table 1. To illustrate the approximate corresponding wavelength range for each DN channel, one photon is assumed to be completely absorbed by one pixel in the conversion. In reality, a highly energetic photon has a finite but very low chance to deposit its energy in more than one pixel, and multiple photons may deposit their energy in one pixel during the 10-sec CCD integration. These possible photon pileup phenomena are beyond the scope of this paper and will be addressed in a future paper with the goal of determining the spectral irradiance by identifying and extracting individual photon events. The majority of the analysis presented in this paper treats the SAM CCD as collectors of incoming photons, deals with contamination statistically, and provides full-disk broadband irradiance. Since the approaches presented in this paper operate on integrals of all the DN bins, photon pileups do not affect the first-order conversion from total DN to broadband irradiance because only the total incident energy is relevant.

According to the WHI quiet-Sun spectrum, about 73% of the solar irradiance shortward of 7 nm lies between 2–7 nm (channels 14 and 15) where SAM is less sensitive. The fraction drops quickly to insignificant toward shorter wavelengths due to lower solar flux. Yet a significant amount of energy is released from shortward of 2 nm during solar flares.

Figure 5 illustrate the relative variability of the signal derived from the difference of the one-hour IA and UA DN histograms (similar to the ones shown in Figure 4) in each SAM channel for year 2011. The fifteen SAM channels are presented along with the GOES 0.1–0.8 nm data and the 10.7 cm radio flux (F10.7) [Tapping, 1989, 2013] as the two are common proxies for solar activities. Since the F10.7 index correlates well with the XPS 0.1–7 nm measurements [Caspi et al., 2015], it should correlate with the overall SAM broadband irradiance as well. Channels are shifted in the Y axis so that their temporal differences are distinguishable. The variability of each data set is evaluated against its own year-long intensity level and presented in the same scale. While the absolute magnitudes of the light curves do not represent much physical information to be compared among channels, the light curves show how differently or similarly channels record the solar irradiance and the dynamic range of the variability at the particular wavelength bands throughout the year. It is clearly observable in Figure 5 that the GOES light curve shows higher range of variation than the longer-wavelength half of the SAM channels as the instrument is designed to be sensitive to the shortest SXR wavelengths. Meanwhile the shorter-wavelength half of the SAM channels present variability comparable to or even greater than that of the GOES measurements. Even when the Sun is quiet in 2010 (not shown here), relatively high variability could occur as a result of contamination that is equivalent to the photons at 0.01–7 nm, such as particles. In the presented scale, the variability of the F10.7 index and lower-energy channels is hardly notable compared to the level of variability of higher-energy channels, which is greater by several orders of magnitude.

Figure 6 shows the correlation of these channels with the GOES XRS A (0.05–0.4 nm) and XRS B (0.1–0.8 nm) channels, the F10.7 index, and the raw SAM broadband irradiance itself. Other than in 2010 (solid), correlation coefficients of the year-long observations show similar trend in 2011 (dot), 2012 (dash), and 2013 (dash dot). Correlation with the GOES channels is high for the shorter wavelengths (higher energy channels) and decreases toward longer wavelengths (lower energy channels). On the other hand, correlation with the F10.7 index increases toward longer wavelength channels. At low solar activities in 2010, the SAM channels are contaminated and overwhelmed by the space environment, resulting in unrealistic correlation curves with the GOES channels (highly variable in SAM but low in GOES) but similar ones with the F10.7 index. Shaded areas indicate the overlapping wavelength ranges with GOES XRS-A (orange) and XRS-B (purple) (the overlapped wavelength between the two in dark red) and out-of-band wavelength range (green) with the assumption that the energy of one photon is completely absorbed by one pixel. Similar correlation analysis is performed exclusively during high solar activities, which is defined as the GOES SXR level above $10^{-6}$ W/m$^2$ at 0.1–0.8 nm (NOAA C-class flares and greater) or above $10^{-7}$ W/m$^2$ in 0.05–0.4 nm, and results are shown in Figure 7. As the GOES XRS is designed to be highly sensitive to solar X-ray flux, the exclusion of data at lower solar activity level improves the correlation. Due to the lack of solar activities, the 2010 data are completely excluded by the criteria and are absent in the figures. Figures 6 and 7 indicate: 1) high correlation between channels 1–8 and the GOES irradiance; 2) high correlation between channels 10–15 with F10.7; and 3) decreasing correlation with broadband irradiance toward longer wavelengths. In other words, the variability of the shorter-wavelength channels dominates that of the broadband irradiance. As shown in Figure 6, the correlation coefficients of the 2010 observations have similar trend with F10.7 as those of the other active years but deviations are present in the correlation with the GOES measurements. On the other hand, the enhancement observed in most of the SAM channels in 2010 likely comes from other contamination sources and are especially prominent in the shorter-wavelength channels. From Figure 5–7, clearly particles and SXR photons reside at similar wavelengths and a deliberate approach has to be taken to separate their contributions. We therefore conclude that the shorter wavelength channels are representative of higher energy solar SXR irradiance or in some cases particles (when the Sun is quiet), while the longer wavelength channels are representative of the lower energy SXR irradiance. This shows there is spectral information in the SAM data, but our present task is to demonstrate that reasonable irradiances can be obtained.

## 4 One-component Method

We now determine irradiance values from the histograms. We will start with a simplified, single parameter approach. Though the best knowledge of non-photon DN is subtracted previously, particle contamination remains in all DN bins as the result of SDO's encounter with the radiation belt. Regardless of the solar activity levels, the enhancement of the received signal due to contaminating particles is close to 10 times more variable than the solar irradiance. This effect is more significant at the higher-energy end of the SAM observing band where solar variability is high. Only during flares does solar irradiance at these high-DN bins rise above the contamination level, as shown in the bottom panel of Figure 4. In the *one-component* analysis method, the raw broadband irradiance, $I_r$, converted from total DN by Equation 2 is assumed to have a combination of contributions from solar irradiance, $I_s$, and particle contamination, $s_p$, as given by Equation 3.

$$I_r = I_{s,1-comp} + s_{p,1-comp} \qquad (3)$$

The terms with subscript, *1-comp*, are used to be distinguished from the later-introduced two-component terms. The 1 or 2 refers to the number of components describing solar irradiance as opposed to contamination. The raw irradiance is the broadband signal obtained by integrating all the pixels above 15 DN except the saturated ones (saturation occurs above 16383 DN). Channels 1–6 at the highest DN/energy have low correlation (< 0.2) with the F10.7 index as well as with GOES at low solar activities (Figure 6), but these channels are not completely quiet during the quiet time owing to the contamination. They are therefore selected to represent the variability of the particle term. Equation 3 is rewritten into Equation 4.

$$I_{BB} = I_{s,1-comp} + A \cdot I_{CH1-6} \qquad (4)$$

The left-hand-side term, noted BB for broadband, is essentially the summation of all the fifteen channels and is the broadband irradiance at 0.01–7 nm. The coefficient, $A$, is a multiplier to represent the solar irradiance variability as characterized by the integrated measurements of channels 1–6, $I_{CH1-6}$. It is a constant and not measureable, but can be estimated. Given the raw broadband and narrowband measurements, each value of $A$ corresponds to a new estimate of $I_{s,1-comp}$, as Equation 4. The best estimate for $A$ is determined by finding the value of $A$ that gives the maximal correlation between the resulting solar term, $I_{s,1-comp}$, and the F10.7 index. Since F10.7 is a daily index, daily averages of broadband and integrated channels over certain period of time are taken to form $I_{BB}$ and $I_{CH1-6}$. Data studied include dates from May 15, 2010 to Dec 31, 2013, providing 1327 daily and 31848 hourly data points. Out of these four years, the Sun is most active in 2011 and quietest in 2010. Estimates of coefficient $A$ are obtained for all individual years with year-long data as well as for the entire four-year period. The comparison among one-year and four-year observations is shown in Figure 8. In the left panel, the one-component estimate of $A$ derived from four-year data (thick dashed line) is 1.52 and the corresponding correlation between solar component, $I_s$, and F10.7 is 0.86. We interpret the high correlation below.

## 5 Two-component Method

Solar irradiance is known to have different degrees of variability at different wavelengths as the solar flux originates at various regions in the solar atmosphere: factors of hundreds in hard X-ray, tens in SXR, and ~2 in EUV [Woods et al., 2004]. Solar models and proxies are often built acknowledging these differences in variability. For instance, the flare component is modeled separately and added to the daily background irradiance in the Flare Irradiance Spectral Model [Chamberlin et al., 2007, 2008] and a hot component (flare) is separated from the quiet-Sun component (background) in the broadband measurements to construct a proxy for GOES XRS [Hock et al., 2013]. Therefore, the assumption of one single solar component in the SAM measurements as in Equation 3 varying with F10.7 is likely not sophisticated enough though it is shown to remove some particle contamination as a reasonable first step toward obtaining cleaner broadband irradiance from SAM images. A quiet $I_q$ and an active $I_a$ component forming the solar term, $I_{s,2\text{-}comp}$ are introduced to represent the variability of quiet and active solar irradiance respectively as given in Equation 5.

$$I_r = I_{s,2-comp} + s_{p,2-comp} = I_q + I_a + s_{p,2-comp} \qquad (5)$$

Channels 1–6 still represent the particle term here as in the one-component method. In the process of building key criteria the GOES 0.1–0.8 nm is the main data set that will be compared to and based on for flaring condition since it is the channel NOAA currently uses to classify flares. The shortest-wavelength channels (1–6) within the narrow wavelength range of 0.05 nm have similarly high correlation with the GOES irradiance likely because they belong to similar groups of hot coronal emissions. High correlation with GOES is the reason why channels 6–8 are selected to characterize the variability of the active component in solar irradiance and substitute for $I_a$ at the right side of Equation 5 with a coefficient *B*. Channels are marked with *'p'* (particle) and *'a'* (active) in Table 1 to indicate their roles in the analysis. Similarly, in the *two-component* method Equation 5 is rewritten into Equation 6 with terms substituted by broadband and integrated irradiance of certain channels.

$$I_{BB} = I_q + B \cdot I_{CH_{6-8}} + A \cdot I_{CH_{1-6}} \tag{6}$$

The inclusion of channel 6 in both active and particle terms allows it to contribute as a particle channel but reserve its accountability for solar photons during flares. A 2D search grid is formed to find the optimal combination of *A* and *B* which result in the maximum correlation between $I_q$ and the F10.7 index. The two-component approach gives *A* a value of 1.08, *B* a value of 0.86, and the highest correlation of 0.88. The middle and right panels of Figure 8 shows the values of *A* and *B* found for all the years. With the introduction of *B*, the value of *A* obtained from all the cases is smaller than the one-component cases. A higher value of *B* tends to result in a greater drop in the value of A in the corresponding case as the contribution from the mutual channel is weighted more in one term than the other but it is not a linear relation.

Both one-component and two-component methods estimate the amount of particle contribution to be removed by obtaining maximum correlation between the solar components (the quiet component in two-component case) with the F10.7 index and reserving their products with the coefficients *A* and *B*. The combination of a higher value of B and a lower value of A suggests a lower contribution from particles in the raw broadband irradiance. With only one term representing the solar contribution in the received signal, the search of A can be biased toward either the particle or the active terms. Thus, the one-component estimate of irradiance, $I_{s,1\text{-}comp}$, consists of both quiet and active parts of the true solar irradiance. On the other hand, the two-component method produces a quiet component, $I_q$, and an active component, $I_a$, which together construct the estimate of broadband irradiance, $I_{s,2\text{-}comp}$. It is worth noting here that $I_{s,2\text{-}comp}$ is not necessarily equal to $I_{s,1\text{-}comp}$ because of different values of $A$ found in these two approaches. Panel (a) in Figure 9 compares the results of the two approaches and shows that the resulting solar irradiance, $I_{s,1\text{-}comp}$, is not as quiet as the quiet component, $I_q$, obtained using two-component method. The slope of the dashed line in the plot is unity and points lying along that line imply equal values of $I_{s,2\text{-}comp}$ and $I_{s,1\text{-}comp}$. The benefits of including an active term in the two-component approach are: 1) it takes back certain portion of the raw signal that is misidentified as particle contribution by the one-component approach and preserves the highly variable portion of the irradiance during flares; 2) it also results in a quieter component, $I_q$, that is closer to the particle-free condition than $I_{s,1\text{-}comp}$. On the other hand, the one-component irradiance, $I_{s,1\text{-}comp}$, contains both quiet and active parts of the solar irradiance though it does not fully capture as high irradiance on the active days nor reach as low at quiet conditions. The normalized difference, $D$, is defined in Equation 7 to quantify the differences of these two estimates of the broadband irradiance.

$$D = \frac{(I_{s,2-comp} - I_{s,1-comp})}{I_{s,1-comp}} = \frac{(I_q + I_a - I_{s,1-comp})}{I_{s,1-comp}} \tag{7}$$

In Figure 9 panel (b), this difference of $I_{s,2\text{-}comp}$ from $I_{s,1\text{-}comp}$ shows that solar irradiance can be either over-estimated by the two-component method or under-estimated by the one-component method at low solar activities (F10.7 < 100). Compared with the GOES level, variability of $I_{s,1\text{-}comp}$ scatters and even falls back down to lower than 1 at high solar activity level while it is not expected to as shown in Figure 9 panel (c). On the other hand, in Figure 9 panel (d) variability of $I_{s,2\text{-}comp}$ is high in some cases even with no apparent X-ray activities. The power-law fit as given in Equation 8 provides a linear fit: a = 1198.86 and b = 0.50 in panel (c) and a = 389.63 and b = 0.42 in panel (d).

$$y = a \cdot x^b \tag{8}$$

This reveals the possible issue of underestimation by the one-component method at high solar activities (GOES > $10^{-6}$) and overestimation by the two-component method at low solar activities (GOES < $10^{-6}$). Therefore, we conclude that one-component estimate of solar irradiance could be adopted during lower solar activities and two-component estimate during higher solar activities.

## 6 A Hybrid Method

A hybrid approach determines the coefficients, $A$ or $A$ and $B$, appropriate to the levels of the solar activity for which either method will be used. The one-component coefficient is used on lower-activity days and the two-component coefficients for higher-activity days. Criteria for estimating the solar activity threshold separating the two approaches need to be established. The relationship of several quantities and the GOES X-ray irradiance are therefore examined. These include: a) one-component estimate, $I_{s,1\text{-}comp}$, b) quiet component, $I_q$, from two-component method, c) difference between $I_{s,1\text{-}comp}$ and $I_q$, the components upon which the correlation coefficients are calculated, d) daily mean (Mean) of the active component, $I_a$, e) standard deviation (Stddev) of the active component, and f) degree of variation defined as the ratio of standard deviation to mean of the active component. In Figure 10, four quadrants in panels (a), (b), (d), and (e) show the effects of different thresholds. Thresholds (dashed lines) of the examined quantities are set to capture the C-class flares (at least $10^{-6}$ W·m$^{-2}$ observed in the GOES 1–8 Å channel) and above in the first quadrant, where the examined quantities is as indicative as the GOES irradiance and the two-component coefficients should be applied to the observations of those flaring days. Data points falling in the fourth quadrant represent the active days that are not recognized. Those falling in the second quadrant are misidentified as the active days while GOES shows lower X-ray irradiance. A careful selection of the Y-axis threshold minimizes possible false alarms in each case. All of the quantities that are examined except the variability of $I_a$ generally serve as good indicators for solar activity level and with slight differences each misses less than 5 flaring days. Table 2 provides the population fractions in the quadrants for each examined quantity. Daily GOES irradiance exceeds $10^{-6}$ W·m$^{-2}$ about 22% out of the 1,467 days included in the analysis. While the threshold set for the quiet components shown in (a) sets apart most of those days, more higher-activity days are failed to be captured when GOES irradiance is between $3\times10^{-6}$ and $10^{-5}$ W·m$^{-2}$ and rising the threshold does not improve the situation. Quantities in panels (b) and (d) have more than three times of the chance misidentifying quiet days as active (quadrant II) than that in panel (e). The ratio of population in quadrant II to (I+II) indicates the probability of a false alarm which happens at the same frequency as the ratio of population in quadrant II to (II+III) during quiet days. The ratio of quadrant I to (I+IV) provides the fraction of the active days actually identified if one solely judges by these quantities. Given the discussion above regarding panel (a), the criterion based on standard deviation of the active component is the best among all selected to be utilized in the process. It provides the highest population in quadrant I, the lowest II/(I+II) and II/(II+III) ratios but the highest I/(I+IV) ratio. When the standard deviation of $I_a$ is greater than $5\times10^{-4}$ W·m$^{-2}$, a day is considered active and the two-component coefficients are employed. Otherwise, the one-component estimate is provided.

## 7 Broadband Irradiance

Four and half years of the broadband derived from the SAM images (hybrid approach) are compared to the ESP irradiance in Figure 11. ESP, also on EVE, is an expanded version of SOHO/SEM. It is a non-focusing broadband spectrograph. Its QD channel produces SXR broadband irradiance at nearly identical wavelengths to SAM. Hourly and daily average of irradiance is shown in panel (a). The green squares indicate the days considered as active by the criteria and the two-component coefficients are applied. The diamonds are color-coded to indicate the strongest flare observed on the particular days: C class in blue, M class in orange, and above X class in red. The dashed lines indicate the 25% difference between the SAM and ESP irradiance where most of the days fall into. It can be observed that the flare class does not necessarily promise higher daily solar irradiance and daily SXR irradiance at 0.01–7 nm does not necessarily correlate with the GOES flare class, which is defined by the peak value of its one-minute measurements within one day span. The scatter plot in panel (b) shows that the ratio of SAM to ESP irradiance is not a function of ESP irradiance. The use of the 2010–2013 data as an entire data set implicitly emphasizes more the variability of active component than on the particle contribution (one quiet year versus three active years). Therefore, the estimated SAM irradiance is higher (at most a factor of 2) than the ESP value on quiet days in 2010. This effect is also seen from the difference between the coefficients obtained for the individual years and all four years in Figure 8. Overall, the SAM/ESP ratio from 2010 to 2014 is close to one. Of all the 31,848 hours studied, the SAM/ESP ratio has a mean of 1.07 and a standard deviation of 0.30. Blue shades beneath the black curve in panel (c) mark the days considered active by the procedure. Clearly the criteria based on the active component has successfully identified the flare days and the procedure properly applies the better set of coefficients to estimate the irradiance. The resulting data sets of SAM broadband irradiance from 2010 to 2014 presented here are available at http://aim.ece.vt.edu/sam/. Data sets with customized date ranges can be made available upon request. This is the entirety of the SAM dataset as MEGS-A encountered a power anomaly of the CCD electronics on May 26, 2014. There can be no observations past that date.

A history of broadband solar SXR measurements are presented in Figure 12. Measurements from SNOE (1998–2003), TIME/SEE (2002–present), SDO/EVE/SAM (2010–present), and SDO/EVE/ESP (2010–present) are shown together in panel (a). Figure 12 panel (b) shows a close-up view from the beginning of the SDO mission to the present. During the second half of 2012, the Sun turned quiet and no major solar events took place. The solar irradiance thus shows clear modulation of the 27-day rotation. It is not unfamiliar that discrepancy of about a factor of two lies among the measurements [BenMoussa et al., 2013; Feng et al., 1989; Solomon et al., 2001] and we suspect that the differences come from the instrumental bandpass differences. Further investigation will be pursued to address this issue and in particular obtaining SXR spectra to better interpret broadband data is essential. SAM can help with this Meanwhile, SAM serves as an addition for further intra-instrument comparison and validation.

## 8 Resolving the Sources of Solar SXR Irradiance

The analysis and validation in previous sections have established the fidelity of deriving broadband irradiance from the whole solar disk from the SAM images. SAM's imaging enables further investigation into spatially resolved features. On February 15, 2011, GOES recorded an X2.3 flare erupting from AR 11158 at the west limb of the Sun. Though the X2.3 flare dominated the X-ray irradiance, minor flares also set off from AR 11161 at the other limb on the same day and contributed to the total irradiance measured by GOES. However, the GOES XRS channels only provide total irradiance and does not provide enough information of the irradiance from each of the active features appearing on the disk at the same time. Current ionospheric and thermospheric models do not require knowledge of which part of the Earth-facing Sun contributing to the observed irradiance. Yet, it is of great interest to the heliophysics community to acquire both spatially and temporal resolved information so that better understanding of the mechanisms behind solar features can be learned, especially when an active Sun has multiple activities that overlap temporally. ESP is unable to provide detailed spatial information about solar SXR features either.

We present an algorithm to determine active areas and obtain irradiance from them based on the high-cadence SAM images. The algorithm finds the brightest features from the SAM images, labels them, and records locations in the SAM coordinates. We obtain the two most active and persistent features in terms of their brightness by performing the algorithm on all the SAM images from February 15, 2011. We create an observation mask around each active feature, obtain the DN histograms and sub-channels, and apply the hybrid technique to derive irradiance at 0.01–7 nm at these two locations. Each of the observation masks is a circular area of radius of 5 pixels, which is about 6% of the solar disk. The preliminary results are shown in Figure 13. The top panel shows that the broadband SXR irradiance derived from two partial ARs from SAM images agrees well with the event observed by other instruments. AR1 (magenta) is located at the center of the flaring AR 11158 where the X2.3 flare erupted. Located at the other limb of the solar disk, AR2 (green) is a part of the AR 11161 where the minor flares occurred. Gaps in the light curves occur when the brightness in the mask is lower than the detectable signal-to-contamination ratio. SAM irradiance presents higher variability than the whole-disk (WD) irradiance from ESP (black). The difference observed between the SAM and ESP curves is mainly contributed by the flaring area outside of

AR1 and the other ~90% of the solar disk. Thus SAM, unlike ESP, is isolating the irradiance from the active region. The GOES XRS channels recorded the event at 0.1–0.8 nm (red) and 0.05–0.4 nm (blue) and their measurements are shown in the bottom panel for comparison. The results demonstrate clearly the advantage of the SAM images for temporal and spatial SXR irradiance.

## 9 Uncertainties

Several sources contribute to the uncertainties in the proposed approaches. The DN cut-off between two histogram bands implicitly assumes that the corresponding DN at both ends of the range of interest is contributed by one photon while using the reference spectrum to determine the scaling factors. As two broad bands are selected to calculate one narrowband channel to mitigate the conversion issue discussed in Section 3, the effect is insignificant as the difference among scaling factors is less than a fraction of one tenth percent from the average for the channels used in the analysis. The major uncertainties come from obtaining values of $A$ and $B$ through correlating one solar component with the F10.7 index in the process of particle contamination removal. They can be estimated by performing the same analysis on each year's data, which maximally changes the value of A by 20% and 50% in the one-component case and the value of B by 55%. The highest deviation in the values of $A$ and $B$ obtained from the one-year data from those obtained from the four-year data occur when the one-component method is performed on the most active year (2011) and the two-component method is performed on the quietest year (2010). This is expected as it is pointed out that the two-component method provides better estimate at the high solar activity levels and the one-component method at the low solar activity levels. Overall, the changes of the values of A and B can vary the irradiance product by 30% and 20% respectively in the one-component and two-component cases. The estimate of these uncertainties is compromised to 27% with the introduction of the hybrid method. The rest of the uncertainty sources including thermal noise and dark counts total about 3 DN at each pixel, which contributes less than one ten thousandth to the broadband irradiance.

## 10 Summary and Conclusions

By treating an array of CCD pixels as a collective photon detector we have shown that it is possible to determine the broadband solar SXR irradiance with the SAM images. Two slices of the SAM images, marked as IA and UA, are used in the analysis for the sunlit and non-sunlit areas of the image. The difference between the IA and UA histograms is integrated and converted to raw broadband signal. The one-component method uses a single solar term to represent the solar contribution in the raw signal while the two-component method uses a quiet and an active solar terms. Fifteen SAM channels are defined and subsets of them are selected as the particle or flare indicators. The coefficients associated with the particle and solar terms are estimated by searching for the highest correlation between the relatively quiet component and F10.7. The one-component estimate of the irradiance does a better job removing the particle contribution at lower solar activities while the two-component estimate performs better at higher solar activities. A criterion based on the level of the extracted active solar component is determined as an indicator to help select either the one-component or two-component irradiance according to the estimated solar activity level. The resulting parameters and criteria from the first four years of data are then applied to the 2014 data and no measurements from other instruments are required further on. The comparison of the 2014 SAM and ESP irradiance shows good agreement within 25% under all solar conditions.

Though SAM and ESP are fundamentally different types of instruments, we have shown that the full-disk broadband SXR irradiance derived from the SAM images agree with the ESP quad-diode measurements within 25%. It is clearly that SAM as an imager is also a valid solar irradiance monitor at 0.01–7 nm. The latency for the data production is one day and the cadence of the broadband product adopting the methods provided in this paper can be as short as 10 seconds.

A preliminary demonstration of SAM's spatial capability is carried out on a day in 2011 when several flares occurred. Broadband irradiance is retrieved locally from two separate bright areas, each of which belongs to an AR that has flared. The size of the two areas combined are about 10% of the solar disk and the light curves derived from them show reasonable agreement with the full-disk ESP measurements. The AR detection algorithm has been fully applied to all the data sets and preliminarily detects up to two ARs at a time. Further validation and improvement are currently undergone and the results will be presented together with the retrieve spectral information in the follow-on paper. With the validation of broadband irradiance derived from the SAM images, we conclude that SAM is indeed a valid solar irradiance monitor at 0.01–7 nm and is a step toward the spectral irradiance. SAM's capability to provide unprecedented high-cadence spatially-resolved information is beneficial for future regional studies. Given the capabilities of SAM, we will continue to retrieve solar SXR spectral irradiance.

## Acknowlegements

This work is funded by funding for EVE science analysis under the SDO mission, NASA contract NAS5-02140 and through a grant from the principal investigator, Tom Woods, at the University of Colorado to Virginia Tech. Amir Caspi was partially funded by NASA grants NNX15AK26G and NNX14AH54G. We thank Brentha Thurairajah, Justin Yonker, Justin Carstens, and Karthik Venkataramani for helpful comments. We are very grateful for many helpful comments and mentorship from Darrel Judge who led EVE ESP channel. We acknowledge the profound impact he has had on our work and our community. Data from 2010 to 2014 are available at http://aim.ece.vt.edu/sam/. Data sets with customized date ranges can be made available upon request. Spatially-resolved broadband irradiances for ARs can be made available upon request; a complete database will be developed and published as part of a future work.

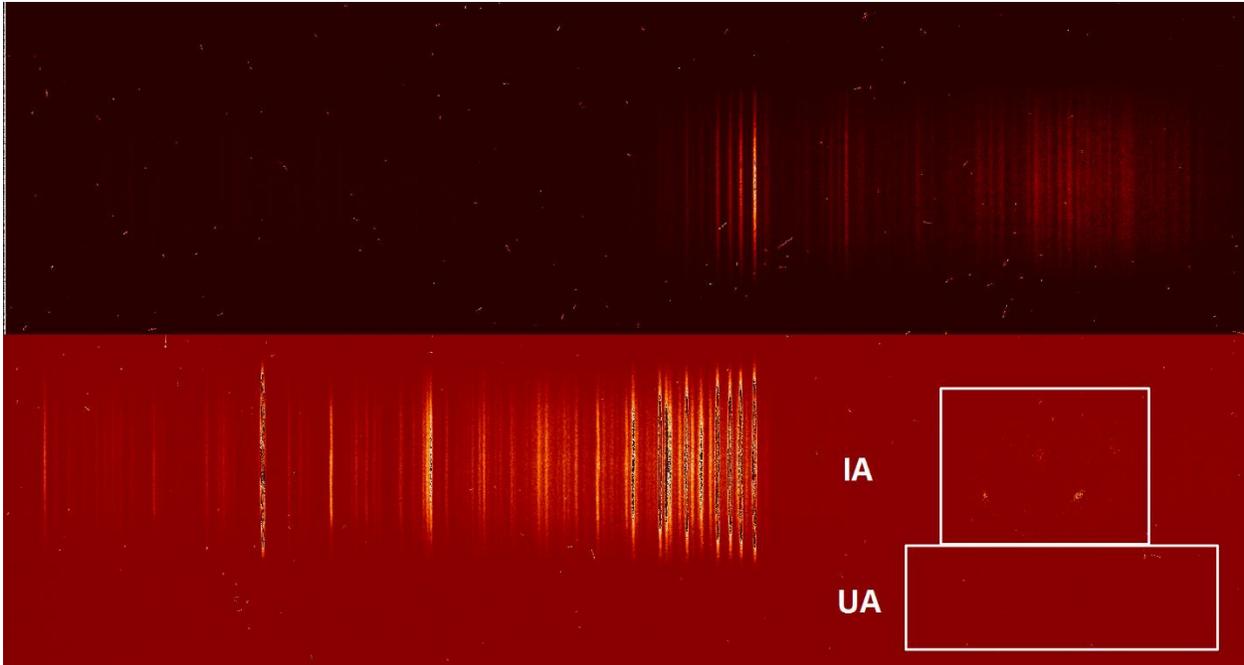

Figure 1. SAM projects solar disk in soft X-ray onto a corner of EVE MEGS-A image at 10-second cadence. The projection is enclosed by an area of 320x240 pixels on the CCD image, referred as IA, where the limb and two active regions are visible. A narrow strip of non-sunlit area, UA, below IA is chosen to represent the particle environment at all times for calibration. The size of UA is 480x160 pixels.

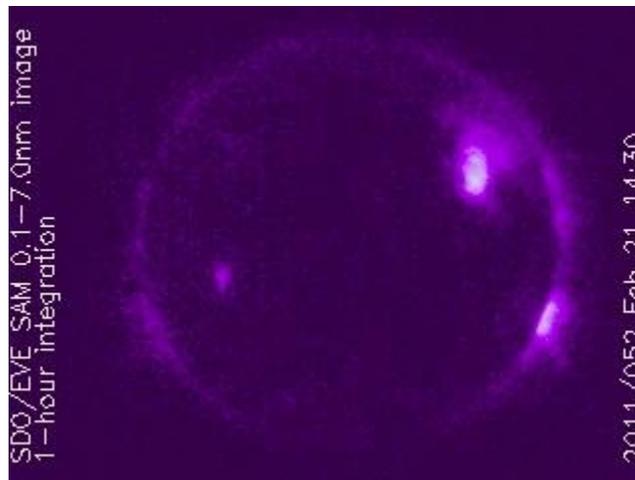

Figure 2. A one-hour integration SAM image [the LASP website] shows the clear limb atmosphere and active regions.

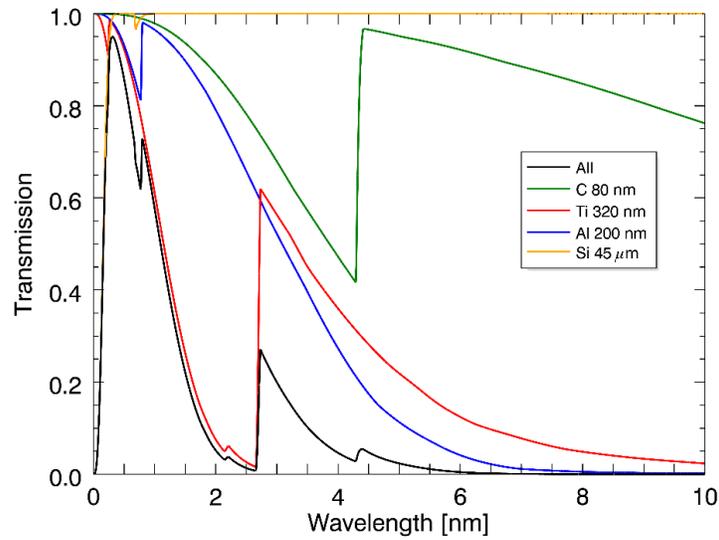

Figure 3. The transmission function of SAM.

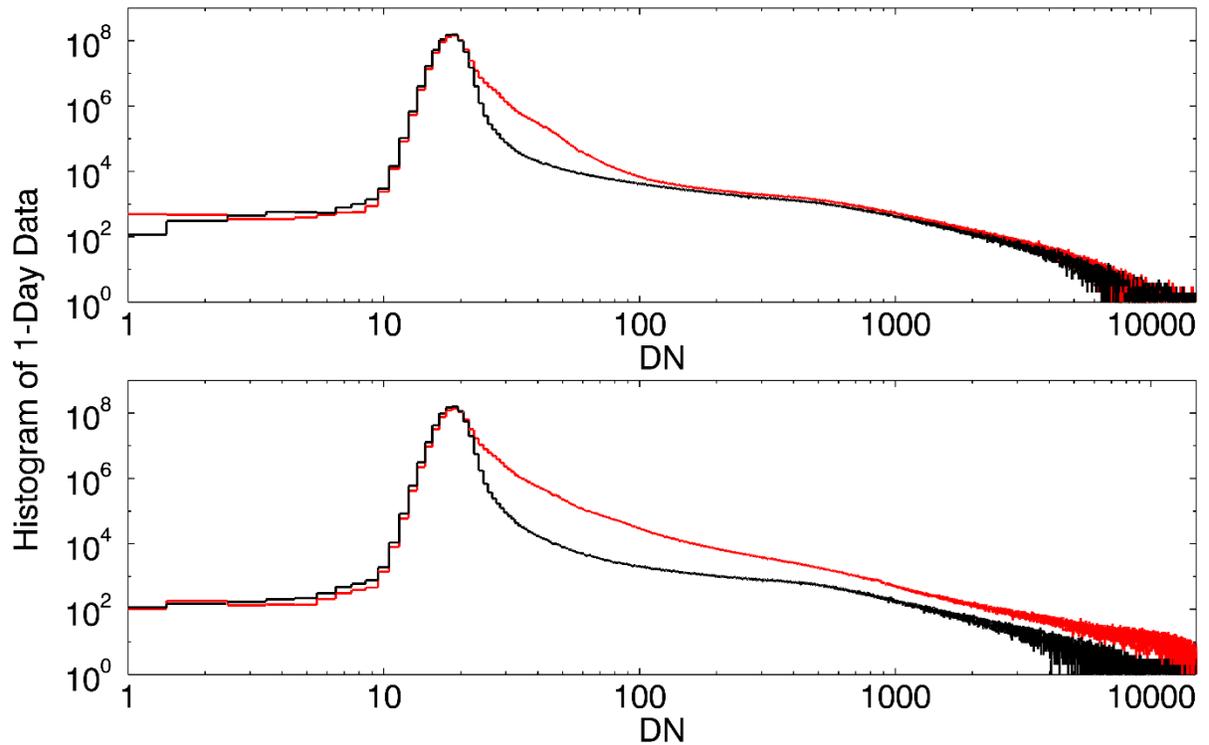

Figure 4. Examples of DN histograms of daily IA (red) and UA (black) images. Each panel contains information of over 8,000 10-second images. The top panel is representative of a typical day of low to medium solar activity and the bottom panel is an extreme case where a strong flare occurred and the IA and UA curves are apart all the way through the high end of the DN distribution.

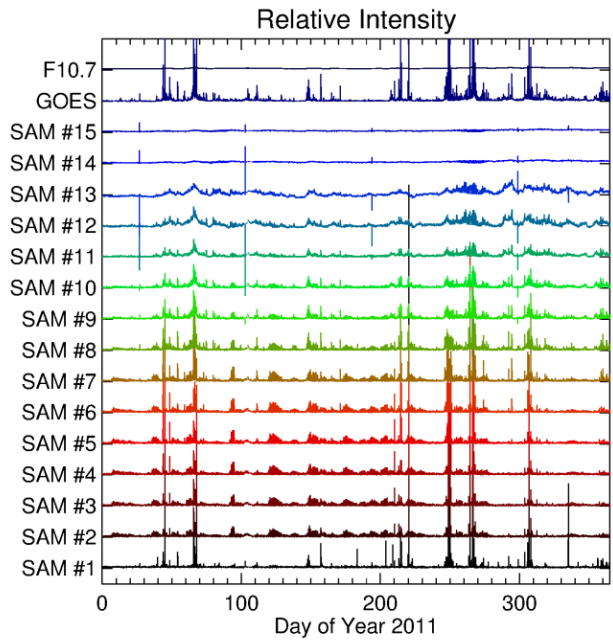

Figure 5. Relative intensity of the light curve at each channel for year 2011. Channels #1 to #15 are the fifteen SAM bands as given in Table 1. The GOES 1–8 Å irradiance and the F10.7 index are also presented. The measurements are shifted in the Y axis so that the differences in the evolution among channels in the year are distinguishable. The variability of each channel is evaluated against its own year-long intensity level. While the absolute magnitudes of the light curves do not represent much physical information to be compared among channels, the light curves show how differently or similarly channels record the solar irradiance and the dynamic range of the variability at the particular wavelength bands throughout the year. In the presented scale, the variability of the F10.7 index (maximally at a scale of a factor of ~2) and lower-energy channels is hardly notable compared to the level of variability of higher-energy channels, which is greater by several orders of magnitude.

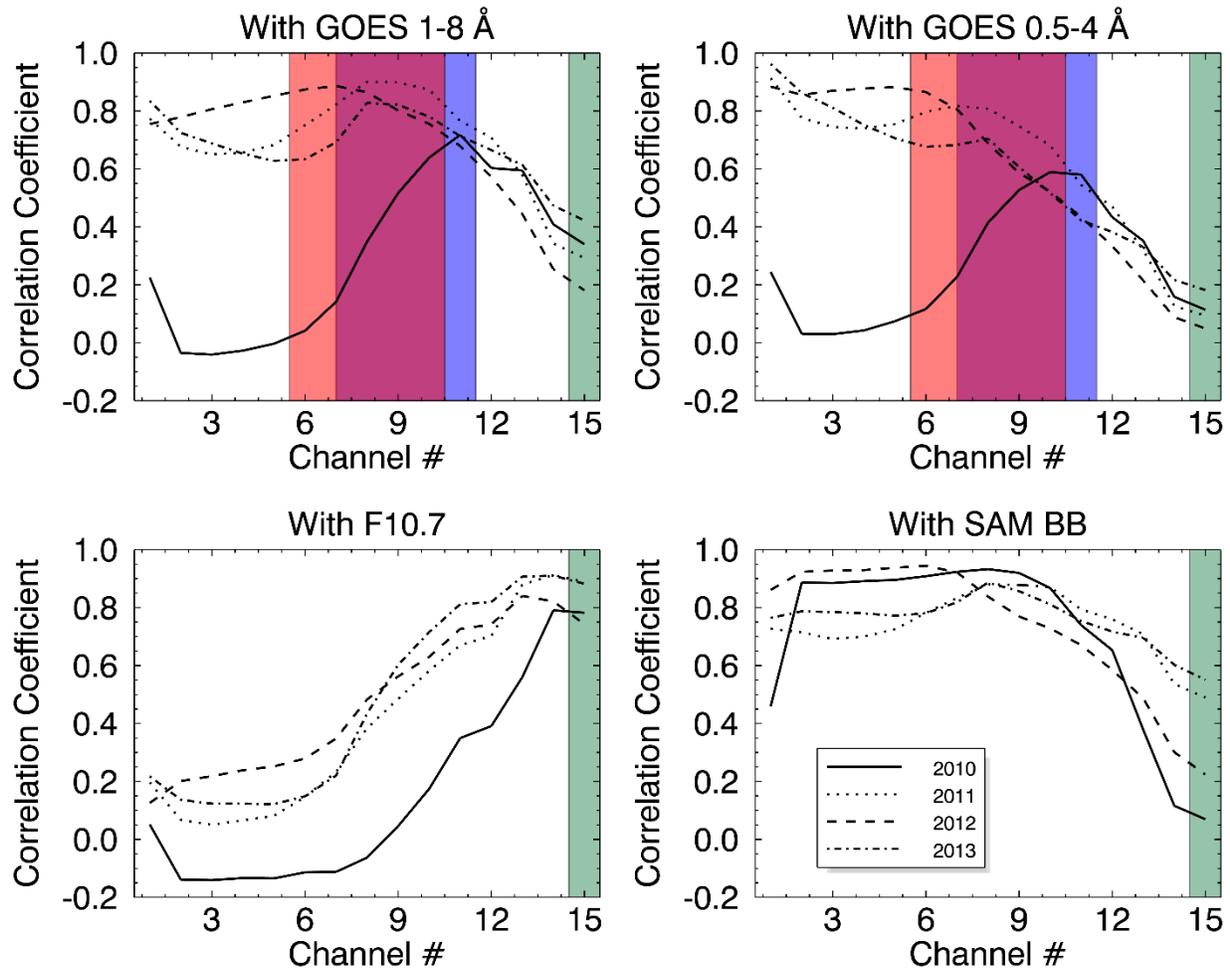

Figure 6. Correlation between the SAM channels and GOES 1–8 Å (top left) and 0.5–1 Å (top right) channels, the F10.7 index (bottom left), and the SAM broadband irradiance (bottom right). Other than in 2010 (solid), correlation coefficients of the year-long observations show similar trend in 2011 (dot), 2012 (dash), and 2013 (dash dot). Correlation with the GOES channels is high at the shorter wavelengths and decreases toward longer wavelengths. On the other hand, correlation with the F10.7 index increases toward longer wavelengths. At low solar activities in 2010, the SAM channels are contaminated by the space environment, resulting in unrealistic correlation curves with the GOES channels but similar ones with the F10.7 index. Shaded areas indicate the overlapping wavelength ranges with GOES 0.5–1 Å (transparent red) and 1–8 Å (transparent red) and out-of-band wavelength range (green) with the assumption that energy of one photon is totally absorbed by one pixel.

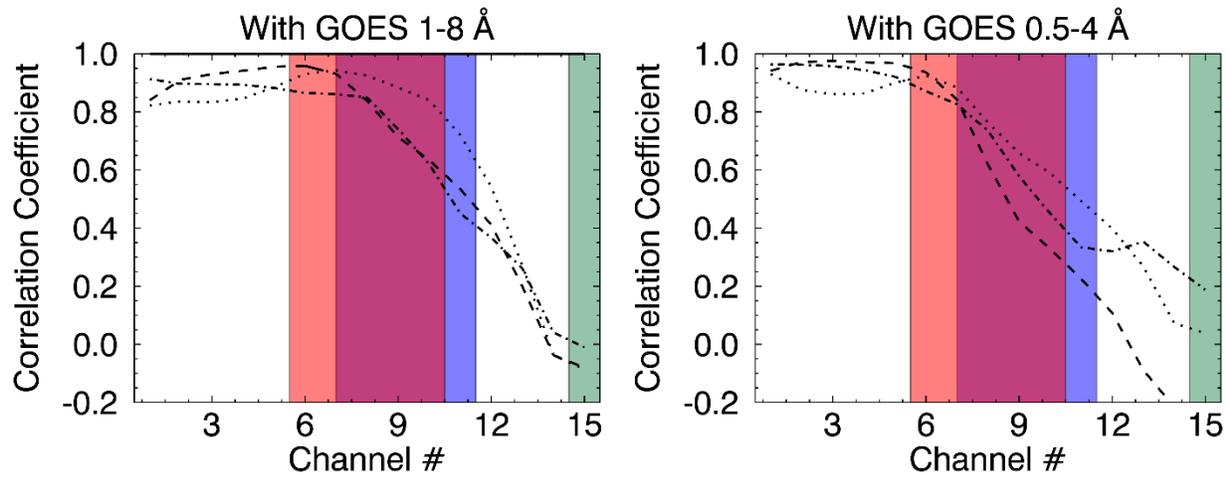

Figure 7. Correlation between the SAM channels and the GOES channels at high solar activity level, which is defined as the GOES SXR level above $10^{-6}$ W/m$^2$ at 1–8 Å or above $10^{-7}$ W/m$^2$ in 0.5–4 Å. As the GOES XRS is designed to be highly sensitive to solar X-ray flux, the exclusion of data at lower solar activity level improves the correlation. The 2010 data is entirely excluded by the criteria.

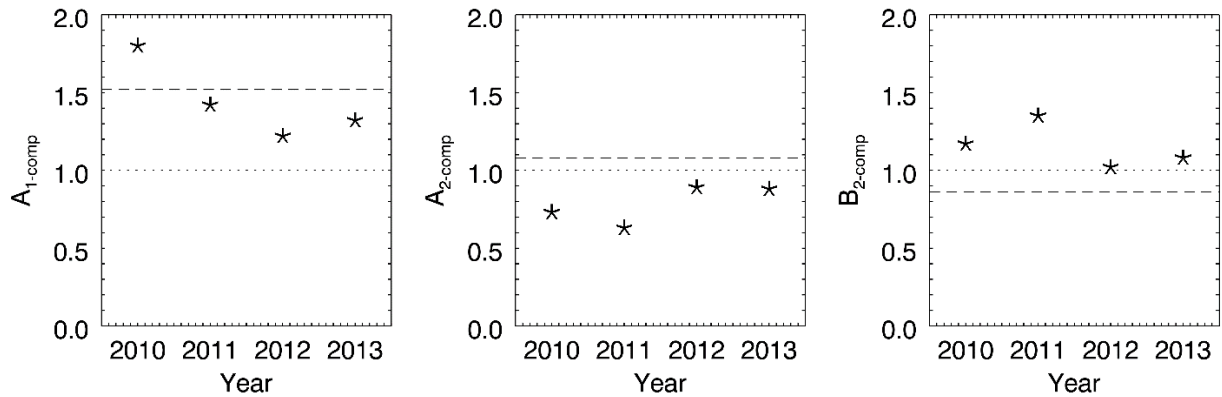

Figure 8. The one-component estimates of coefficient $A$ (left) are greater than unity (thin dash) for all individual years while the two-component ones (middle) are less than unity with the introduction of coefficient $B$ (right). Year 2010 is the only year whose one-component estimate of $A$ is greater than the four-year estimate (thick dash).

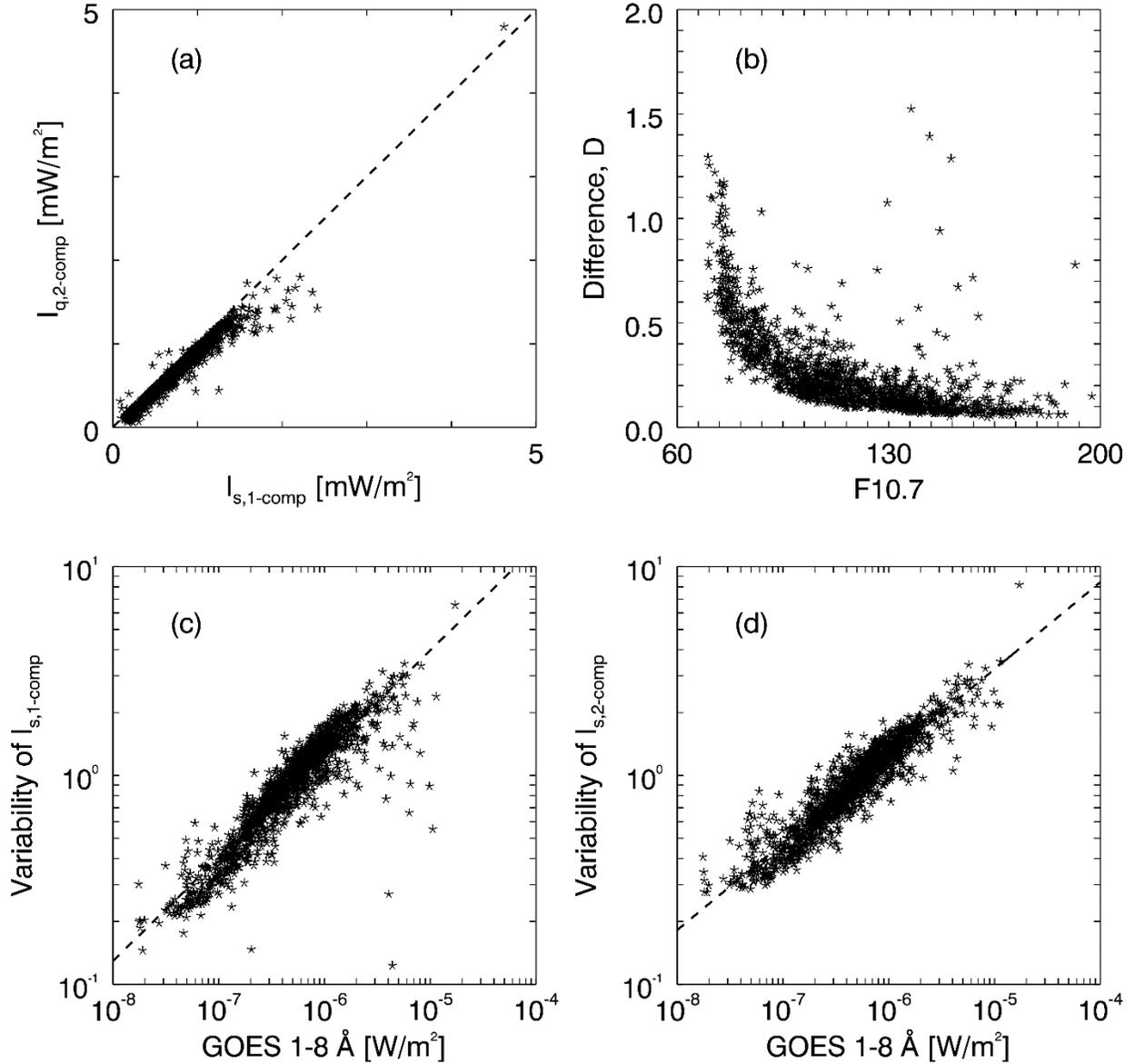

Figure 9. Comparison of products of the one-component and two-component methods. Two-component method provides two solar terms, $I_q$ and $I_a$, whereas one-component method produces only one solar component, $I_s$, which contains partial irradiance of the active Sun and is higher than that of the two-component quiet irradiance as shown in panel (a). In panel (b), the difference between the one-component and two-component estimates is higher at low solar activities (F10.7 < 100). While compared with the GOES level, variability of $I_{s,1\text{-}comp}$ scatters and even falls back down to lower than unity at high solar activity level as shown in panel (c) and variability of $I_{s,2\text{-}comp}$ is higher in a significant number of cases without apparent X-ray activities as shown in panel (d). The dashed lines in panels (c) and (d) are the power-law fitting of the cases as given in Equation 9: A = 1198.86 and B = 0.50 in panel (c); A = 389.63 and B = 0.42 in panel (d).

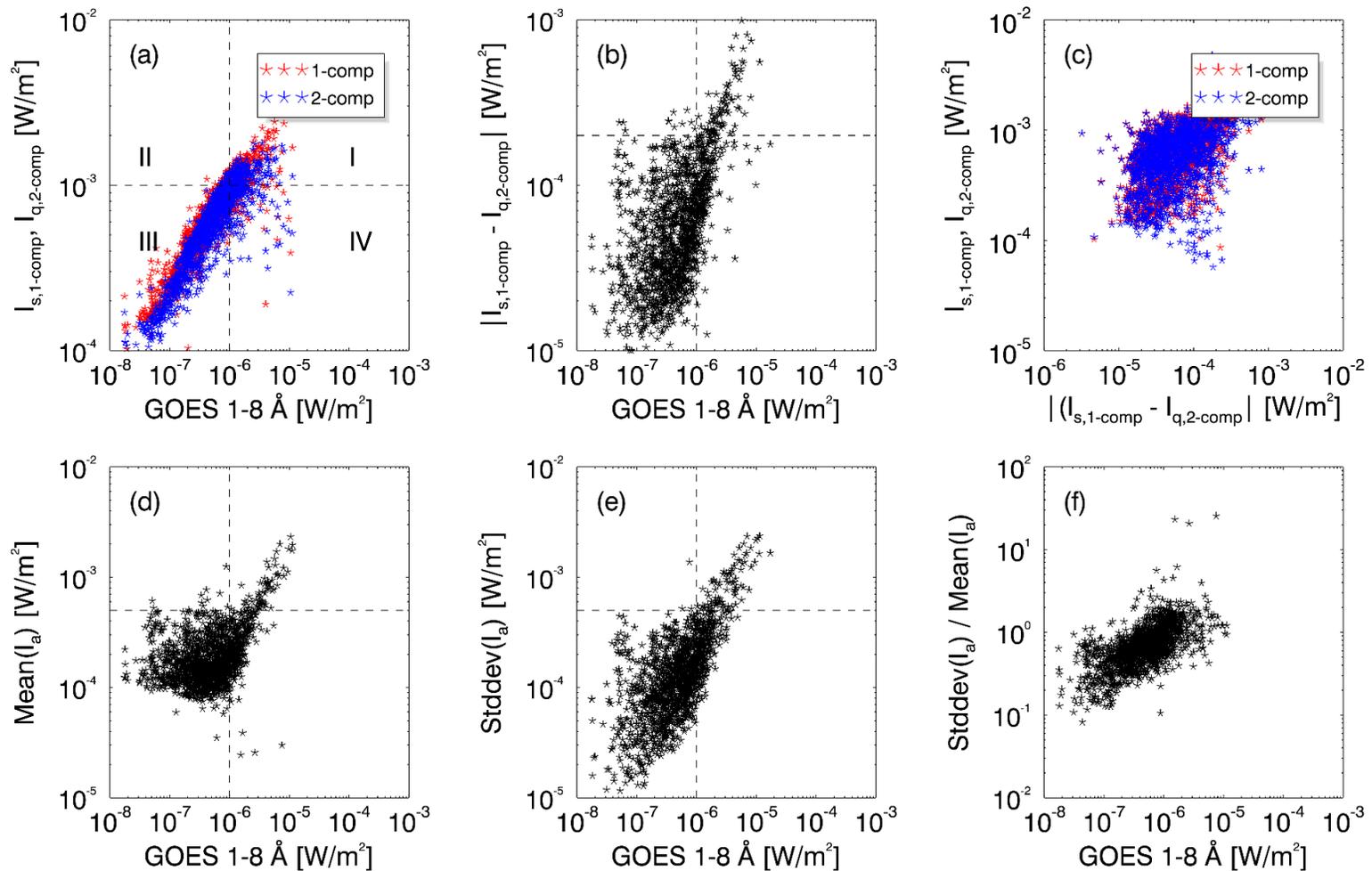

Figure 10. Relationship of several quantities and the GOES XRS irradiance is examined to help determine the criteria for applying proper sets of coefficients. Quantities examined include: extracted solar component, $I_s$, from one-component method (red) and quiet-Sun component, $I_q$, from two-component method (blue) shown in panel (a); difference between the two in panel (b); daily mean of the active component, $I_a$, as in Equation 4 in panel (d); 5) standard deviation of $I_a$ in panel (e), and degree of variation defined as its ratio of standard deviation to mean of in panel (f). The quieter components have little correlation with the difference between each other as shown in panel (c). Four quadrants are defined by two dashed lines in the comparison with the GOES X-ray data. The right-hand side of the vertical line indicates C-class and above flares. The horizontal line defines the quadrant able to capture the flare condition when the two-component coefficients should be applied.

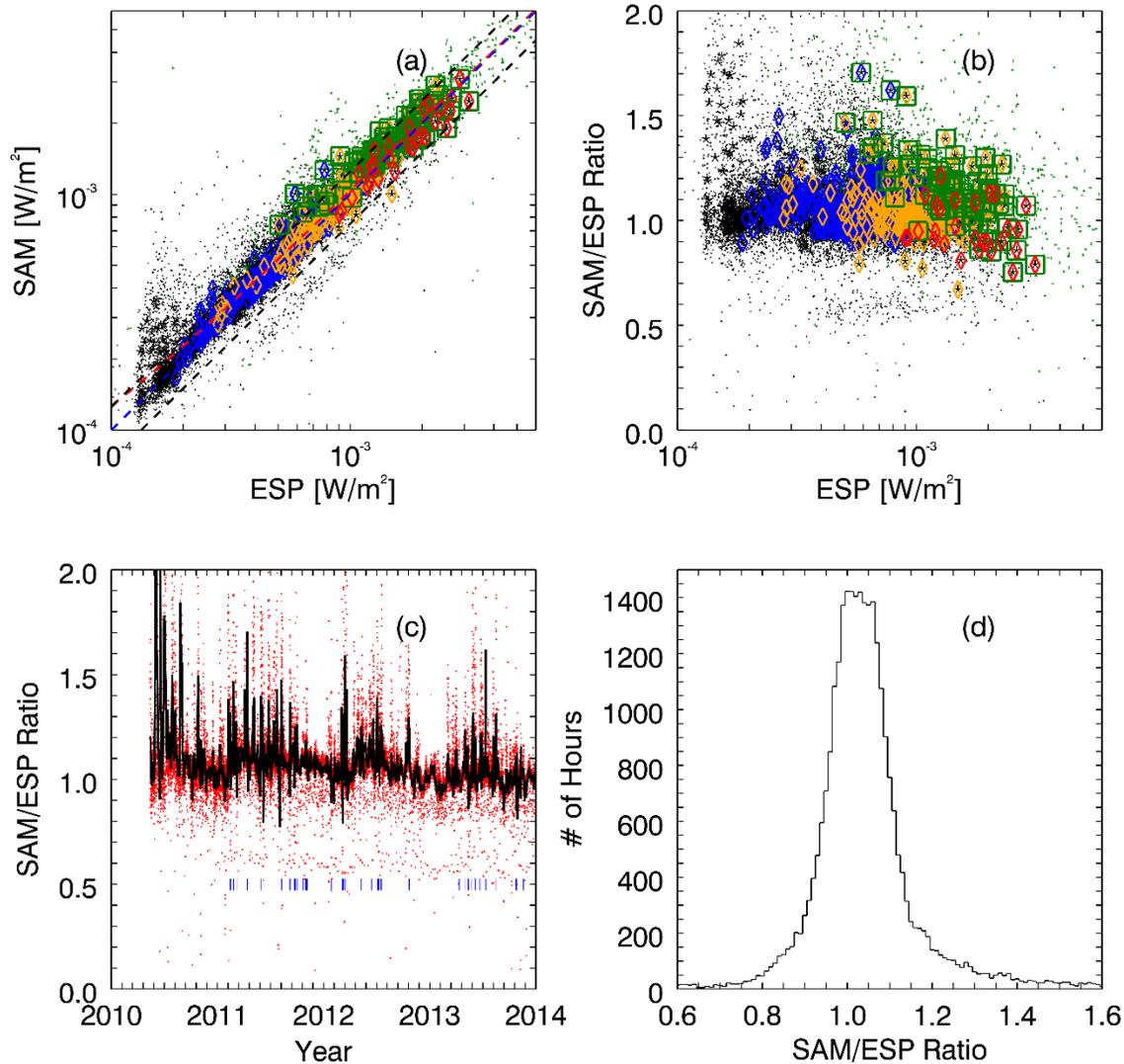

Figure 11. Comparison of broadband irradiance from SAM and ESP is illustrated. (a) Hourly irradiance is shown with dots and daily mean with asterisks. The green squares indicate the days recognized as active and the 2-component coefficients are applied. The diamonds are color-coded to indicate the flare strength on the particular days: C class in blue, M class in orange, and X class in red. The daily soft X-ray irradiance does not necessarily correlate with the GOES flare class, which is defined as the peak value of its one-minute measurements within one day span. Data points are fitted with a line (red), which is close to the X = Y line (blue). The dashed lines indicate the 25% difference between the SAM and ESP irradiance. Green squares indicate the days recognized as active ones by the procedure. Most of the data points reside within the 25% region. The hybrid approach successfully recognizes the active days and the 2-component coefficients are applied to improve the estimated irradiance that would otherwise falls outside of the 25% region. (b) The scatter plot shows that the ratio of the SAM to ESP irradiance is not a function of the ESP irradiance. (c) The hourly (red dots) and daily (black) ratio from 2010 to 2014 is close to unity. When standard deviation of $s_a$ reaches $5 \times 10^{-4}$, a day is considered active and indicated by a blue strip beneath the irradiance curve. (d) Of the 31,848 hours studied, the SAM/ESP ratio has a mean of 1.07 and a standard deviation of 0.30.

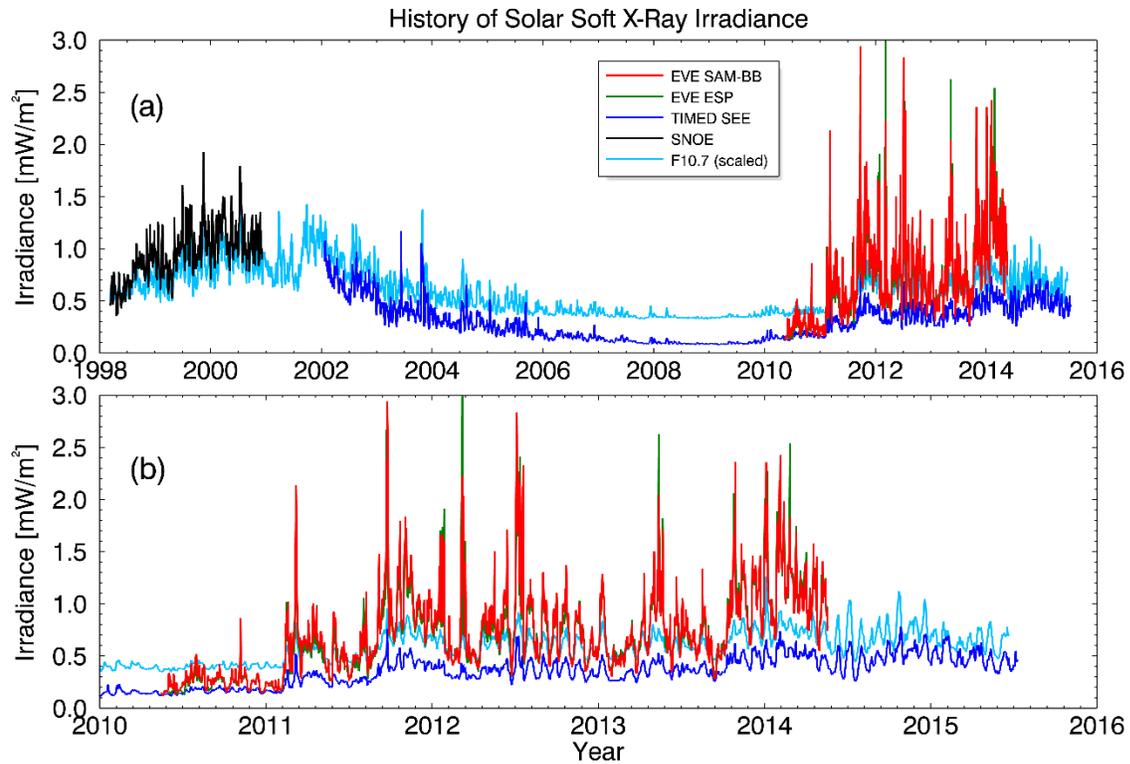

Figure 12. Time series of solar soft X-ray measurements and the scaled F10.7 index from 1998 (a) and from the beginning of the SDO mission (b) to present. During the second half of 2012, the Sun turned quiet and therefore the solar irradiance shows clear modulation of its 27-day rotation.

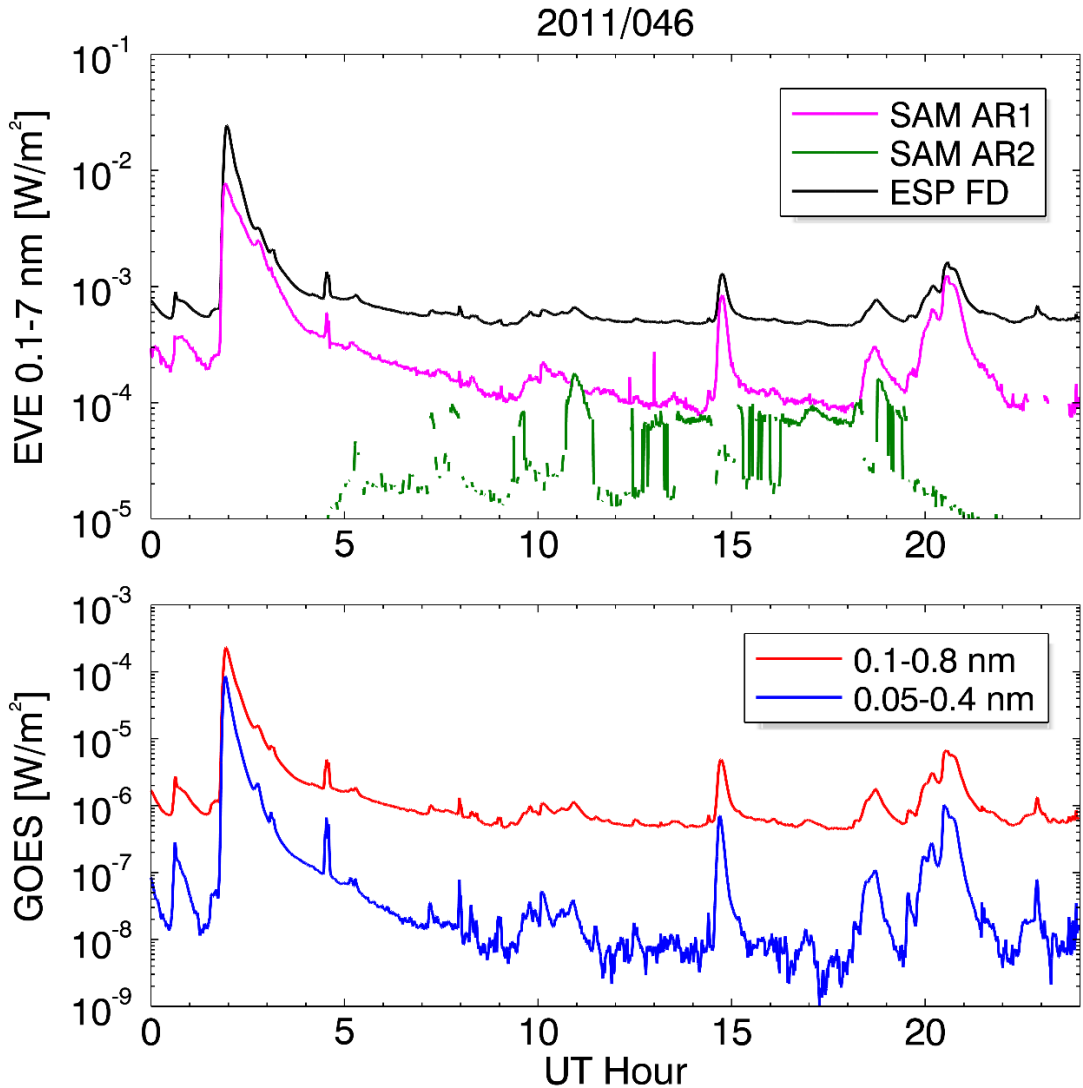

Figure 13. (Top) The broadband SXR irradiance derived from two partial ARs, each of which is about 6% the size of the solar disk, from SAM images agrees very well with the event observed by other instruments. AR1 (magenta) is at the center of the flaring AR 11158 and AR2 (green) is part of the AR 11161, which is located at the other limb of the solar disk. It presents higher variability than the full-disk (FD) irradiance from ESP (black). The other 90+% of the solar disk contributes to most of the difference observed between the two light curves shown here. (Bottom) The GOES XRS channels recorded the event at 0.1–0.8 nm (red) and 0.05–0.4 nm (blue).

Table 1. Definition of narrow bands – upper and lower limits in DN, energy, and wavelength. To minimize the uncertainties caused by assuming energy of one photon is completely absorbed by one pixel, signal strength of a narrow band is determined by the difference of that of two broad bands.

| Band $i$ | DN min | DN max | Energy[a] [keV] From | Energy[a] [keV] To | Wavelength[a] [nm] From | Wavelength[a] [nm] To | % | Note[b] |
|---|---|---|---|---|---|---|---|---|
| 1 | 7000 | 16383 | 62.76 | 146.89 | 8.4e-3 | 2.0e-2 | * | $p$ |
| 2 | 6000 | 6999 | 53.80 | 62.76 | 2.0e-2 | 2.3e-2 | * | $p$ |
| 3 | 5000 | 5999 | 44.83 | 53.80 | 2.3e-2 | 2.8e-2 | * | $p$ |
| 4 | 4000 | 4999 | 35.86 | 44.83 | 2.8e-2 | 3.5e-2 | * | $p$ |
| 5 | 3000 | 3999 | 26.90 | 35.86 | 3.5e-2 | 4.6e-2 | * | $p$ |
| 6 | 2000 | 2999 | 17.93 | 26.90 | 4.6e-2 | 6.9e-2 | * | $p, a$ |
| 7 | 1000 | 1999 | 8.97 | 17.93 | 6.9e-2 | 1.4e-1 | * | $a$ |
| 8 | 500 | 999 | 4.48 | 8.97 | 1.4e-1 | 2.8e-1 | * | $a$ |
| 9 | 400 | 499 | 3.59 | 4.48 | 2.8e-1 | 3.5e-1 | * | |
| 10 | 300 | 399 | 2.69 | 3.59 | 3.5e-1 | 4.6e-1 | ~$10^{-3}$ | |
| 11 | 200 | 299 | 1.79 | 2.69 | 4.6e-1 | 6.9e-1 | 0.1 | |
| 12 | 100 | 199 | 0.90 | 1.79 | 6.9e-1 | 1.4 | 2.1 | |
| 13 | 50 | 99 | 0.45 | 0.90 | 1.38 | 2.77 | 24.6 | |
| 14 | 20 | 49 | 0.18 | 0.45 | 2.77 | 6.91 | 73.0 | |
| 15 | 15 | 19 | 0.13 | 0.18 | 6.91 | 9.22 | | |

[a]The conversion from DN to energy and wavelength is performed under the assumption that energy of a photon is completely absorbed by a pixel.
[b]The channels used to characterize particle term and active component are marked as 'p' and 'a' respectively.
*The fraction is too small to be accurately accounted for due to model constrains.

Table 2. Population in each quadrant and the ratios examined.

| | I | II | III | IV | II/(I+II) | I/(I+IV) | II/(II+III) |
|---|---|---|---|---|---|---|---|
| $s_{1,comp}$ | 16.29% | 3.48% | 73.69% | 5.66% | 17.59% | 71.34% | 4.51% |
| $s_q$ | 11.32% | 1.50% | 75.66% | 10.77% | 11.70% | 49.55% | 1.94% |
| $|s_{1,comp}-s_q|$ | 5.59% | 0.68% | 76.41% | 16.36% | 10.87% | 24.48% | 0.88% |
| Mean of $s_a$ | 4.16% | 0.75% | 76.41% | 17.93% | 15.28% | 18.21% | 0.97% |
| Stddev of $s_a$ | 5.66% | 0.20% | 76.96% | 16.43% | 3.49% | 24.78% | 0.27% |